\documentclass[rnote]{aa}
\usepackage{txfonts}
\usepackage{graphicx}
\begin{document}

\title{The angular sizes of dwarf stars and subgiants}
\subtitle{Non-linear surface brightness relations in $BVR_cI_c$ from interferometry}
\titlerunning{The angular sizes of dwarf stars and subgiants from $BVR_cI_c$ colors}
\authorrunning{P. Kervella \& P. Fouqu\'e}
\author{
P. Kervella\inst{1}
\and
P. Fouqu\'e\inst{2}
}
\offprints{P. Kervella}
\mail{Pierre.Kervella@obspm.fr}
\institute{
LESIA, Observatoire de Paris, CNRS\,UMR\,8109, UPMC, Universit\'e Paris Diderot,
5 Place Jules Janssen, 92195 Meudon, France
\and
LATT, Universit\'e de Toulouse, CNRS, 
14 avenue Edouard Belin, 31400 Toulouse, France
}
\date{Received ; Accepted}
\abstract
{The prediction of stellar angular diameters from broadband photometry plays an important 
role for different applications. In particular, long-baseline interferometry, 
gravitational microlensing, extrasolar planet transits, and many other observing 
techniques require accurate predictions of the angular size of stars. These predictions 
are based on the surface brightness-color (SBC) relations.}
{Our goal is to calibrate general-purpose SBC relations using visible colors, the most
commonly available data for most stars.}
{We compiled the existing long-baseline interferometric observations of nearby dwarf and 
subgiant stars and the corresponding broadband photometry in the Johnson $BV$ and Cousins 
$R_cI_c$ bands. We then adjusted polynomial SBC models to these data.}
{Due to the presence of spectral features that depend on the effective temperature, the 
SBC relations are usually not linear for visible colors. We present polynomial fits
that can be employed with $BVR_cI_c$ based colors to predict the limb-darkened angular 
diameters (i.e. photospheric) of dwarf and subgiant stars with a typical accuracy of 5\%.}
{The derived polynomial relations provide a satisfactory approximation to the observed 
surface brightness of dwarfs and subgiants. For distant stars, the interstellar reddening
should be taken into account.}
\keywords{Stars: fundamental parameters, Techniques: interferometric}

\maketitle

\section{Introduction}
The surface brightness-color (hereafter SBC) relations link the emerging flux per
solid angle unit of a light-emitting body to its color. They express the 
Stefan-Boltzmann relation between total brightness and effective temperature
in measurable quantities such as angular diameter, magnitude and color. It
basically assumes that temperature and bolometric correction can be translated 
into a color measurement.
These relations have many astrophysical applications, such as the estimation of
Cepheid distances (through the Baade-Wesselink method), extrasolar planet
transit studies or the characterization of microlensing sources.
Different colors produce tighter or more dispersed relations, simply because
color also depends on other variables such as gravity or metallicity, at a different
level depending on the adopted photometric bands. It has been shown that a combination
of visible and near-IR bands, such as $(V-K)$, is probably optimal (see, e.g.,
Fouqu\'e \& Gieren~\cite{fouque97}, in the case of Cepheids). However, near-infrared
photometry is not always available, for instance due to catalogue incompleteness or
field crowding of IR atlases such as 2MASS in the Galactic disk.
As a consequence, SBC relations in the visible domain remain very useful.

Although they are relatively similar for stars of different luminosity 
classes, the literature gives appropriate relations for Cepheids (Kervella et al.~\cite{kervella04c}), 
giants (van Belle~\cite{vanbelle99a}, Nordgren et al.~\cite{nordgren02}), M giants
and supergiants (van Belle et al.~\cite{vanbelle99b}, Groenewegen~\cite{groen04}),
and dwarf or subgiant stars (Kervella et al.~\cite{kervella04b}, hereafter K04).
Here we limit our discussion to dwarf stars and subgiants, and revise and extend
K04's work, which was based on Johnson photometric bands and was limited
to linear SBC relations. In the present Note, we calibrate the SBC relations based
on Cousins $R_c$ and $I_c$  photometric bands, of more common use than Johnson's
$R$ and $I$, and we adjust non-linear SBC relations that are a better fit to the
observations than linear laws. In this process, we use exclusively direct interferometric
angular diameter measurements, thus avoiding cross-correlations with previously calibrated
SBC relations.

\section{Interferometric and photometric data \label{ipdata}}

We collected from the literature all the available interferometric
angular diameter measurements of dwarf and subgiant stars. From this
list, we removed the stars for which the angular diameter 
was uncertain by more than 5\%. 
The major sets of new interferometric measurements of dwarf and subgiant stars
since K04 are from Berger et al.~(\cite{berger06}),
Baines et al.~(\cite{baines08}), Boyajian et al.~(\cite{boyajian08})
and Kervella et al.~(\cite{kervella08}),  all four based on measurements obtained with
the CHARA array (ten Brummelaar et al.~\cite{tenbrummelaar05}).  The
only star from Berger et al.~(\cite{berger06}) that has an angular
diameter more precise than 5\%, GJ\,15A, was removed from our list as
it is a flare star, and its photometry is therefore uncertain.
This was also the case for several other M dwarfs.
Most of the candidate stars present variability at a certain level. In order to reject those stars
for which the variability amplitude could be large, we checked their variability status in the GCPD
(Mermilliod et al. \cite{mermilliod97}) and the GCVS (Samus~\cite{samus08}).
All stars listed in ``$\gamma$\,Cas'' or ``UV\,Ceti'' classes have been rejected.
Low amplitude ``$\delta$\,Scuti'', such as \object{$\alpha$ Lyr} or
\object{$\beta$ Leo}, ``BY\,Dra'' like \object{$\epsilon$ Eri},
\object{GJ 570 A}, \object{GJ 699}, \object{61 Cyg A}, 
\object{HR 511}, or possible ``RS\,Cvn'', such as \object{$\delta$ Eri}, have been kept.
As a remark, \object{DY Eri} (a ``UV\,Ceti'' type flare star) is not \object{GJ 166 A}, as stated in the
SIMBAD database, but \object{GJ 166 C}, therefore we kept the former in our list.
We also removed the known very fast rotating stars ($v \sin i \gtrsim 100$\,km.s$^{-1}$), including the pole-on rotator
\object{$\alpha$ Lyr} (Aufdenberg et al.~\cite{aufdenberg06}; Peterson et al.~\cite{peterson06}).
The effective temperature of fast rotating stars is inhomogeneous over their apparent disk, they
can be surrounded by circumstellar material (Kervella \& Domiciano de Souza~\cite{kervella06})
and their photospheres can be spectacularly distorted (Monnier et al.~\cite{monnier07}).
All these phenomena combine to bias their surface brightness compared to normal stars.
Overall, our selection procedure resulted in the removal of about one
third of the available interferometric measurements (Table~\ref{interfdata}).

The limb darkened (LD) angular diameters were available in all cases from the original
publication, except the measurementd of Sirius by Hanbury Brown et al.~(\cite{hanbury74}) and
Davis \& Tango~(\cite{davis86}), for which we applied the LD correction by Claret~(\cite{claret00}),
as discussed in Sect.~5.4 of Kervella et al.~(\cite{kervella04a}).
For Mozurkewich et al.'s~(\cite{mozurkewich03}) measurements we considered the combined LD values
listed by these authors.
In all cases, we chose to keep the original LD values instead of correcting the uniform disk values by
recomputing the limb darkening corrections, because there is a remarkable concensus in the literature
to use the predictions from Kurucz's atmosphere models, approximated using Claret's laws
(Claret et al.~\cite{claret95}; Claret~\cite{claret98}; Claret~\cite{claret00})
to estimate the brightness distribution over the stellar disk.
In fact, all the listed angular diameters were computed using these LD models,
except the recent measurement of $\alpha$\,Cen~B published by Bigot et al.~(\cite{bigot06})
who employed 3D hydrodynamical models. Even in this case, the derived angular diameter
was less than 1$\sigma$ away from the 1D Kurucz model result.
Claret~(\cite{claret08}) recently showed that theoretical atmosphere models do not perfectly
reproduce the observed stellar intensity profiles from transiting exoplanets and eclipsing binaries.
However, these discrepancies are negligible for our purpose, considering the star-to-star
residual dispersion with respect to the adjusted empirical laws (Sect.~\ref{modelfit}).

The starting point for the photometry listed in Table~\ref{photomdata} is the Hipparcos Input
Catalogue (\cite{hipparcos93}) as retrieved from SIMBAD, which lists $V$ and $B-V$, and the Hipparcos and
Tycho Catalogues (\cite{hipparcos97}), which list $V-I$. Notes in these catalogues give the origin of the
listed values. When original photometry in Cousins system exists for $V-R_c$ and $V-I_c$, it was checked from 
the on-line version of the Lausanne General Catalogue of Photometric Data (\cite{mermilliod97}), 
averaged over original sources according to their number of measurements, and finally combined to the
$V$ band Hipparcos photometry to get $R_c$ and $I_c$ magnitudes. More details are given in the Notes of Table~\ref{photomdata}.

\begin{table}
\caption{Interferometric data. The error bars are listed in small characters as exponents.}
\label{interfdata}
\begin{tabular}{lcrrrrcrc}
\hline
Star & Spect. & $\lambda$ & $\theta_{\rm UD}$\,(mas) & $\theta_{\rm LD}$\,(mas) & Ref. \\
\hline
\noalign{\smallskip}
\object{$\alpha$ CMa A} & A1V & B & $ 5.600 ^{ 0.070 } $ & $ 5.896 ^{ 0.074 } $ &  1 \\
\object{$\alpha$ CMa A} & A1V & B & $ 5.630 ^{ 0.080 } $ & $ 5.928 ^{ 0.084 } $ &  2 \\
\object{$\alpha$ CMa A} & A1V & K & $ 5.936 ^{ 0.016 } $ & $ 6.039 ^{ 0.019 } $ &  3 \\
\object{$\alpha$ CMa A} & A1V & V & $  $ & $ 5.993 ^{ 0.108 } $ &  4 \\
\object{$\alpha$ PsA} & A3V & K & $ 2.197 ^{ 0.023 } $ & $ 2.228 ^{ 0.023 } $ &  5 \\
\object{$\beta$ Leo} & A3V & K & $ 1.429 ^{ 0.027 } $ & $ 1.449 ^{ 0.027 } $ &  5 \\
\object{94 Cet} & F8V & K & $ 0.774 ^{ 0.026 } $ & $ 0.788 ^{ 0.026 } $ &  6 \\
\object{$\alpha$ CMi A} & F5IV-V & V & $ 5.190 ^{ 0.040 } $ & $ 5.430 ^{ 0.070 } $ &  7 \\
\object{$\alpha$ CMi A} & F5IV-V & V & $  $ & $ 5.446 ^{ 0.054 } $ &  4 \\
\object{$\alpha$ CMi A} & F5IV-V & K & $ 5.376 ^{ 0.047 } $ & $ 5.448 ^{ 0.053 } $ &  8 \\
\object{$\tau$ Boo} & F7V & K & $ 0.771 ^{ 0.015 } $ & $ 0.786 ^{ 0.016 } $ &  6 \\
\object{$\upsilon$ And} & F8V & K & $ 1.091 ^{ 0.009 } $ & $ 1.114 ^{ 0.009 } $ &  6 \\
\object{$\eta$ Boo} & G0IV & V & $ 2.170 ^{ 0.060 } $ & $ 2.280 ^{ 0.070 } $ &  7 \\
\object{$\eta$ Boo} & G0IV & V & $  $ & $ 2.269 ^{ 0.025 } $ &  4 \\
\object{$\eta$ Boo} & G0IV & K & $     $ & $ 2.200 ^{ 0.031 } $ &  9 \\
\object{$\mu$ Her} & G5IV & V & $  $ & $ 1.953 ^{ 0.039 } $ &  4 \\
\object{$\zeta$ Her A} & G0IV & V & $  $ & $ 2.367 ^{ 0.051 } $ &  4 \\
\object{$\zeta$ Her A} & G0IV & V & $ 2.370 ^{ 0.080 } $ & $ 2.490 ^{ 0.090 } $ &  7 \\
\object{$\alpha$ Cen A} & G2V & K & $ 8.314 ^{ 0.016 } $ & $ 8.511 ^{ 0.020 } $ &  10 \\
\object{Sun} & G2V &  & $     $ & $ 1919260^{10} $ &   \\
\object{70 Vir} & G5V & K & $ 0.986 ^{ 0.023 } $ & $ 1.009 ^{ 0.024 } $ &  6 \\
\object{$\mu$ Cas A} & G5Vp & K & $ 0.951 ^{ 0.009 } $ & $ 0.973 ^{ 0.009 } $ &  11 \\
\object{HR 7670} & G6IV & K & $ 0.682 ^{ 0.019 } $ & $ 0.698 ^{ 0.019 } $ &  6 \\
\object{55 Cnc} & G8V & K & $ 0.834 ^{ 0.024 } $ & $ 0.854 ^{ 0.024 } $ &  6 \\
\object{$\beta$ Aql} & G8IV & V & $ 2.070 ^{ 0.090 } $ & $ 2.180 ^{ 0.090 } $ &  12 \\
\object{$\tau$ Cet} & G8V & K & $ 2.032 ^{ 0.031 } $ & $ 2.078 ^{ 0.031 } $ &  5 \\
\object{54 Psc} & K0V & K & $ 0.773 ^{ 0.026 } $ & $ 0.790 ^{ 0.027 } $ &  6 \\
\object{$\sigma$ Dra} & K0V & K & $ 1.224 ^{ 0.011 } $ & $ 1.254 ^{ 0.012 } $ &  11 \\
\object{HR 511} & K0V & K & $ 0.747 ^{ 0.021 } $ & $ 0.763 ^{ 0.021 } $ &  11 \\
\object{$\delta$ Eri} & K0IV & K & $     $ & $ 2.394 ^{ 0.029 } $ &  9 \\
\object{$\eta$ Cep} & K0IV & V & $ 2.510 ^{ 0.040 } $ & $ 2.650 ^{ 0.040 } $ &  12 \\
\object{$\alpha$ Cen B} & K1V & K & $ 5.881 ^{ 0.021 } $ & $ 6.000 ^{ 0.021 } $ &  13 \\
\object{GJ 166A} & K1V & K & $ 1.600 ^{ 0.060 } $ & $ 1.650 ^{ 0.060 } $ &  14 \\
\object{$\epsilon$ Eri} & K2V & K & $ 2.093 ^{ 0.029 } $ & $ 2.148 ^{ 0.029 } $ &  5 \\
\object{GJ 570A} & K4V & K & $ 1.190 ^{ 0.030 } $ & $ 1.230 ^{ 0.030 } $ &  14 \\
\object{GJ 845} & K4.5V & K & $ 1.840 ^{ 0.020 } $ & $ 1.890 ^{ 0.020 } $ &  14 \\
\object{61 Cyg A} & K5V & K & $     $ & $ 1.775 ^{ 0.013 } $ &  15 \\
\object{61 Cyg B} & K7V & K & $     $ & $ 1.581 ^{ 0.022 } $ &  15 \\
\object{GJ 380} & K7V & HK & $ 1.268 ^{ 0.040 } $ & $ 1.175 ^{ 0.040 } $ &  16 \\
\object{GJ 887} & M0.5V & K & $ 1.366 ^{ 0.040 } $ & $ 1.388 ^{ 0.040 } $ &  17 \\
\object{GJ 411} & M1.5V & HK & $ 1.413 ^{ 0.030 } $ & $ 1.464 ^{ 0.030 } $ &  16 \\
\object{GJ 699} & M4Ve & HK & $ 0.987 ^{ 0.040 } $ & $ 1.026 ^{ 0.040 } $ &  16 \\
\noalign{\smallskip}
\hline
\noalign{\smallskip}
\end{tabular}
References:
(1) Hanbury Brown et al.~\cite{hanbury74};
(2) Davis \& Tango~\cite{davis86}.
(3) Kervella et al.~\cite{kervella03a};
(4) Mozurkewich et al.~\cite{mozurkewich03};
(5) Di Folco et al.~\cite{difolco04};
(6) Baines et al.~\cite{baines08};
(7) Nordgren et al.~\cite{nordgren01};
(8) Kervella et al.~\cite{kervella04a};
(9) Th\'evenin et al.~\cite{thevenin05};
(10) Kervella et al.~\cite{kervella03b};
(11) Boyajian et al.~\cite{boyajian08}.
(12) Nordgren et al.~\cite{nordgren99};
(13) Bigot et al.~\cite{bigot06};
(14) Kervella et al.~\cite{kervella04b};
(15) Kervella et al.~\cite{kervella08};
(16) Lane et al.~\cite{lane01};
(17) S\'egransan et al.~\cite{segransan03};
\end{table}

\begin{table}
\caption{Photometric data. When the uncertainty on the photometric measurements was not available,
we adopted an arbitrary $\pm 0.05$ value.}
\label{photomdata}
\begin{tabular}{lccccc}
\hline
Star & $B$ & $V$ & $R_c$ & $I_c$ & Ref. \\
\hline
\noalign{\smallskip}
\object{$\alpha$\,CMa\,A}  & -$1.43^{0.01}$ & -$1.44 ^{ 0.01 } $ & -$1.43 ^{ 0.05 } $ & -$1.42 ^{ 0.05 } $ & ab \\
\object{$\alpha$\,PsA}  & $ 1.25 ^{ 0.02 } $ & $ 1.17 ^{ 0.02 } $ & $ 1.11 ^{ 0.05 } $ & $ 1.09 ^{ 0.05 } $ & bc \\
\object{$\beta$\,Leo}  & $ 2.23 ^{ 0.01 } $ & $ 2.14 ^{ 0.01 } $ & $     $ & $ 2.04 ^{ 0.05 } $ & H \\
\object{94\,Cet}  & $ 5.65 ^{ 0.01 } $ & $ 5.07 ^{ 0.01 } $ & $ 4.75 ^{ 0.05 } $ & $ 4.44 ^{ 0.05 } $ & c \\
\object{$\alpha$\,CMi\,A}  & $ 0.83 ^{ 0.03 } $ & $ 0.40 ^{ 0.03 } $ & $ 0.15 ^{ 0.06 } $ & -$0.09 ^{ 0.06 } $ & bc \\
\object{$\tau$\,Boo}  & $ 5.00 ^{ 0.03 } $ & $ 4.49 ^{ 0.03 } $ & $     $ & $ 3.98 ^{ 0.04 } $ & C \\
\object{$\upsilon$\,And}  & $ 4.63 ^{ 0.01 } $ & $ 4.09 ^{ 0.01 } $ & $     $ & $ 3.51 ^{ 0.03 } $ & C \\
\object{$\eta$\,Boo}  & $ 3.26 ^{ 0.01 } $ & $ 2.68 ^{ 0.01 } $ & $     $ & $ 2.03 ^{ 0.05 } $ & H \\
\object{$\mu$\,Her}  & $ 4.16 ^{ 0.03 } $ & $ 3.41 ^{ 0.03 } $ & $     $ & $ 2.70 ^{ 0.04 } $ & G \\
\object{$\zeta$\,Her A}  & $ 3.46 ^{ 0.01 } $ & $ 2.81 ^{ 0.01 } $ & $     $ & $ 2.11 ^{ 0.03 } $ & G \\
\object{$\alpha$\,Cen A}  & $ 0.70 ^{ 0.05 } $ & -$0.01 ^{ 0.03 } $ & -$0.37 ^{ 0.06 } $ & -$0.70 ^{ 0.06 } $ & a \\
\object{Sun}  & -$26.11^{0.02}$ & -$26.75^{0.01}$ & -$27.10^{0.01}$ & -$27.44^{0.01}$ & S \\
\object{70\,Vir}  & $ 5.69 ^{ 0.01 } $ & $ 4.98 ^{ 0.01 } $ & $     $ & $ 4.21 ^{ 0.05 } $ & G \\
\object{$\mu$\,Cas\,A}  & $ 5.87 ^{ 0.03 } $ & $ 5.17 ^{ 0.03 } $ & $     $ & $ 4.34 ^{ 0.04 } $ & G \\
\object{HR\,7670}  & $ 6.50 ^{ 0.01 } $ & $ 5.75 ^{ 0.01 } $ & $     $ & $ 4.92 ^{ 0.07 } $ & F \\
\object{55\,Cnc}  & $ 6.83 ^{ 0.02 } $ & $ 5.96 ^{ 0.01 } $ & $     $ & $ 5.13 ^{ 0.07 } $ & F* \\
\object{$\beta$\,Aql}  & $ 4.57 ^{ 0.01 } $ & $ 3.72 ^{ 0.01 } $ & $ 3.26 ^{ 0.05 } $ & $ 2.83 ^{ 0.05 } $ & bc \\
\object{$\tau$\,Cet}  & $ 4.22 ^{ 0.01 } $ & $ 3.50 ^{ 0.01 } $ & $ 3.07 ^{ 0.05 } $ & $ 2.68 ^{ 0.05 } $ & bc \\
\object{54\,Psc}  & $ 6.73 ^{ 0.02 } $ & $ 5.88 ^{ 0.02 } $ & $     $ & $ 5.05 ^{ 0.02 } $ & G \\
\object{$\sigma$\,Dra}  & $ 5.46 ^{ 0.01 } $ & $ 4.68 ^{ 0.01 } $ & $     $ & $ 3.83 ^{ 0.02 } $ & G \\
\object{HR\,511}  & $ 6.43 ^{ 0.01 } $ & $ 5.63 ^{ 0.01 } $ & $     $ & $ 4.82 ^{ 0.05 } $ & G \\
\object{$\delta$\,Eri}  & $ 4.44 ^{ 0.02 } $ & $ 3.53 ^{ 0.01 } $ & $ 3.02 ^{ 0.05 } $ & $ 2.59 ^{ 0.05 } $ & bc \\
\object{$\eta$\,Cep}  & $ 4.34 ^{ 0.01 } $ & $ 3.43 ^{ 0.01 } $ & $     $ & $ 2.49 ^{ 0.04 } $ & G \\
\object{$\alpha$\,Cen B}  & $ 2.25 ^{ 0.04 } $ & $ 1.35 ^{ 0.03 } $ & $ 0.88 ^{ 0.06 } $ & $ 0.47 ^{ 0.06 } $ & a \\
\object{GJ\,166A}  & $ 5.25 ^{ 0.01 } $ & $ 4.43 ^{ 0.01 } $ & $ 3.96 ^{ 0.05 } $ & $ 3.54 ^{ 0.05 } $ & bc \\
\object{$\epsilon$\,Eri}  & $ 4.61 ^{ 0.01 } $ & $ 3.73 ^{ 0.01 } $ & $ 3.22 ^{ 0.05 } $ & $ 2.78 ^{ 0.05 } $ & bc \\
\object{GJ\,570A}  & $ 6.75 ^{ 0.03 } $ & $ 5.72 ^{ 0.03 } $ & $ 5.07 ^{ 0.06 } $ & $ 4.54 ^{ 0.06 } $ & d \\
\object{GJ\,845}  & $ 5.74 ^{ 0.02 } $ & $ 4.69 ^{ 0.01 } $ & $ 4.06 ^{ 0.05 } $ & $ 3.54 ^{ 0.05 } $ & bcd \\
\object{61\,Cyg\,A}  & $ 6.27 ^{ 0.03 } $ & $ 5.20 ^{ 0.03 } $ & $     $ & $ 3.89 ^{ 0.04 } $ & C* \\
\object{61\,Cyg\,B}  & $ 7.41 ^{ 0.03 } $ & $ 6.06 ^{ 0.02 } $ & $     $ & $ 4.50 ^{ 0.04 } $ & C* \\
\object{GJ\,380}  & $ 7.97 ^{ 0.02 } $ & $ 6.60 ^{ 0.01 } $ & $     $ & $ 5.09 ^{ 0.02 } $ & I* \\
\object{GJ\,887}  & $ 8.83 ^{ 0.01 } $ & $ 7.35 ^{ 0.01 } $ & $ 6.38 ^{ 0.05 } $ & $ 5.33 ^{ 0.05 } $ & ce \\
\object{GJ\,411}  & $ 9.00 ^{ 0.02 } $ & $ 7.50 ^{ 0.01 } $ & $ 6.49 ^{ 0.05 } $ & $ 5.35 ^{ 0.05 } $ & d \\
\object{GJ\,699}  & $ 11.11 ^{ 0.03 } $ & $ 9.54 ^{ 0.03 } $ & $ 8.31 ^{ 0.06 } $ & $ 6.76 ^{ 0.06 } $ & acdf \\
\noalign{\smallskip}
\hline
\noalign{\smallskip}
\end{tabular}
References:
(a) Bessell~\cite{bessell90};
(b) Cousins~\cite{cousins80b};
(c) Cousins~\cite{cousins80a};
(d) Celis~\cite{celis86};
(e) Th\'e~\cite{the84};
(f) Laing~\cite{laing89};
(S) Holmberg et al.~\cite{holmberg06};
(C) Hipparcos catalogue, Field H42 = C
(C*) for 61 Cyg A and B, we transformed $(V-I)_j$ from Ducati~(\cite{ducati02}) to $(V-I)_c$ according to Hipparcos precepts when Field H42 = C
(F) Hipparcos catalogue, Field H42 = F
(F*) for 55 Cnc, we converted $(R-I)_c = 0.388 \pm 0.003$ from Taylor~(\cite{taylor03}) to $(V-I)_c$ according to Hipparcos precepts when Field H42 = F
(G) Hipparcos catalogue, Field H42 = G
(H) Hipparcos catalogue, Field H42 = H
(I*) for GJ380, we converted $(B-V)_t$ from Tycho to $(V-I)_c$ according to Hipparcos precepts when Field H42 = I
\end{table}

\section{Surface brightness-color relations \label{modelfit}}

Contrary to visible-infrared colors from K04, the dependence of the zero-magnitude 
limb-darkened angular diameter (ZMLD, defined for $m_\lambda = 0$), as a
function of the color is non linear in the visible. For this reason,
we selected polynomial SBC relation models of the form:
\begin{equation}
\log(\theta_{\rm LD}) = a_0 + a_1\,C + a_2\,C^2 + a_3\,C^3 + a_4\,C^4- 0.2\,m_\lambda
\end{equation}
where $C$ is the color of the star from two dereddened photometric bands among $BVR_cI_c$
(e.g. $B-V$, $V-R_c$,...), $m_\lambda$ the dereddened magnitude of the star
in one of the bands, and $\theta_{\rm LD}$ the limb-darkened angular diameter, measured in
milliarcseconds (mas). Note that this expression is independent of the distance of the star.
For the empirical fit to the observations, we selected the smallest
polynome degree that gave a satisfactory fit of all the data points. This led
us to select a higher degree (four) for the $B-V$ based polynomial fits.

\begin{figure}[ht]
\sidecaption
\includegraphics[width=8.5cm]{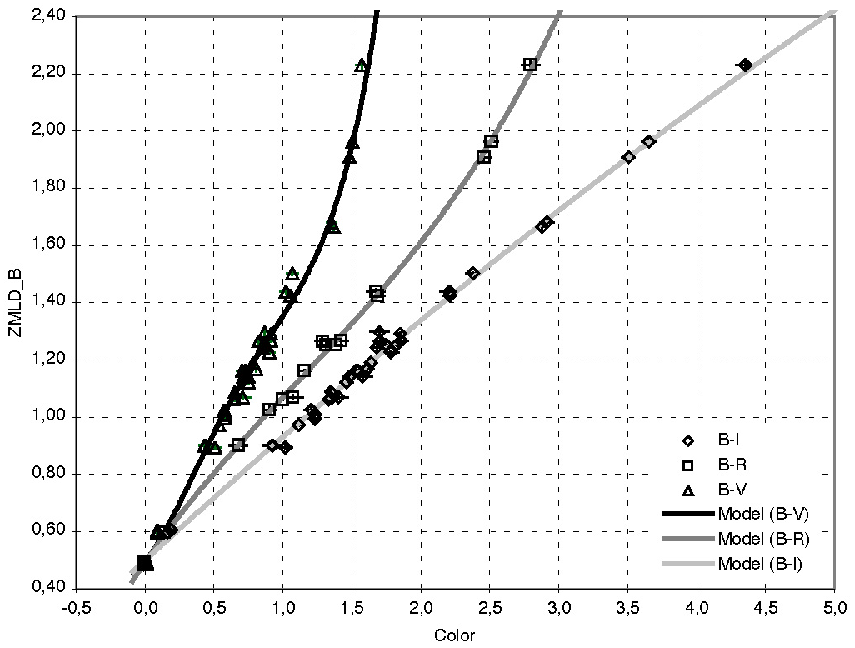}
\includegraphics[width=8.5cm]{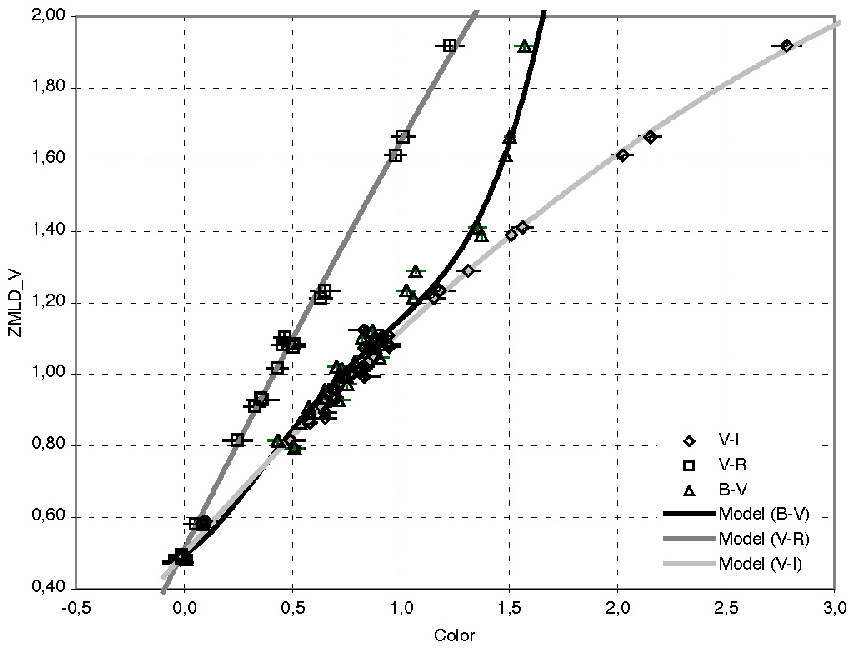}
\caption{Adjusted polynomial relations giving the zero-magnitude limb-darkened 
disk angular diameter (ZMLD$_\lambda$) in the $B$ and $V$
bands as a function of different color combinations. The
corresponding polynomial coefficients are listed in Table~\ref{coefficients}.}
\label{polyfit}
\end{figure}

\begin{figure}[ht]
\sidecaption
\includegraphics[width=8.5cm]{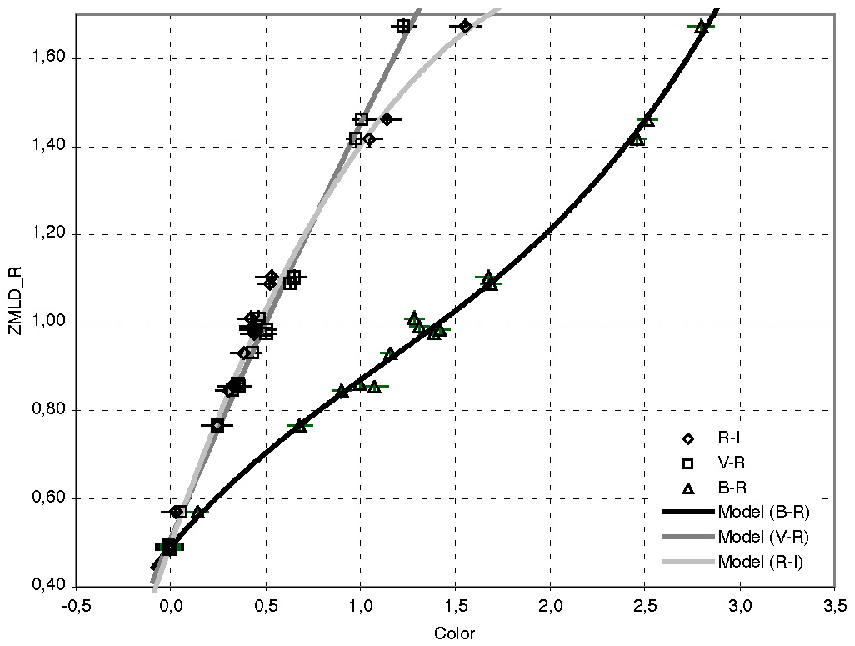}
\includegraphics[width=8.5cm]{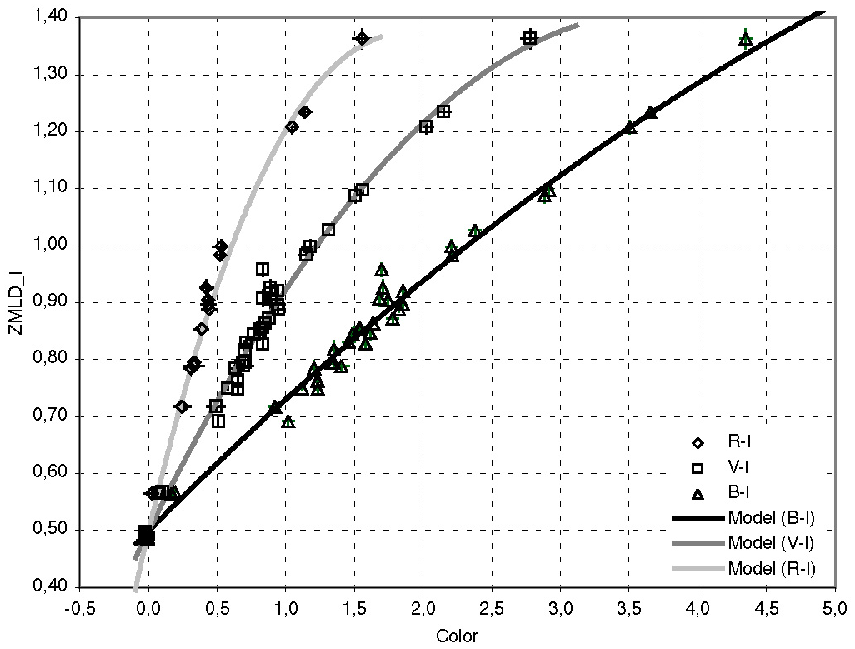}
\caption{Adjusted polynomial relations in the $R_c$ and $I_c$ bands.}
\label{polyfit2}
\end{figure}

We would like to stress that the computed relations are valid only for
a given range of colors, with only little margin for extrapolation.
In particular, the SBC laws undergo
steep variations for red objects. The results of the fits are presented in
Fig.~\ref{polyfit} for $B$ and $V$ and in Fig.~\ref{polyfit2} for $R_c$ and $I_c$.
The polynomial coefficients and the domain of validity of each relation,
are listed in Table~\ref{coefficients}.
The residual dispersions $\sigma(\theta_{\rm LD})/\theta_{\rm LD}$ give the
relative accuracy of the angular diameter predictions for each relation.

\begin{table*}
\caption{Polynomial coefficients of the adjusted surface brightness-color relations:
$\log(\theta_{\rm LD}) = a_0 + a_1\,C + a_2\,C^2 + a_3\,C^3 + a_4\,C^4 - 0.2\,m_\lambda$
($\theta_{\rm LD}$ expressed in milliarcseconds). $N$ is the number of measurements used for the calibration.
The listed $C_{\rm min}$ and $C_{\rm max}$ values correspond to the minimum and 
maximum color index of the stars included in the polynomial fitting procedure. They 
represent the limits outside of which the adjusted relations are unreliable and should 
not be employed.
The standard deviation of the residuals of the fit in log scale is listed in the $\sigma(\log \theta_{\rm LD})$
column, and the corresponding relative uncertainties on the predicted angular
diameter is given in the rightmost column, in percentage.
}
\label{coefficients}
\begin{tabular}{lccrcccccccccc}
\hline
 $\lambda$  &  $C$ & $N$ &  $C_{\rm min}$ & $C_{\rm max}$ & $a_0$ & $a_1$ & $a_2$ & $a_3$  & $a_4$ & $\sigma(\log \theta_{\rm LD})$ & $\sigma(\theta_{\rm LD})/\theta_{\rm LD}$ \\
\hline
\noalign{\smallskip}
$ B $ & $ B-V $ &  42 & 0.01 & 1.57 & 0.4952 & 0.5809 & 1.3259 & -1.7191 & 0.6715 & 0.0331 & 7.9\% \\
$ B $ & $ B-R_c $ &  21 & 0.00 & 2.80 & 0.4922 & 0.6933 & -0.1639 & 0.0486 &  & 0.0172 & 4.0\% \\
$ B $ & $ B-I_c $ &  42 & -0.01 & 4.35 & 0.4996 & 0.4424 & -0.0115 &  &  & 0.0213 & 5.0\% \\
$ V $ & $ B-V $ &  42 & 0.01 & 1.57 & 0.4952 & 0.3809 & 1.3261 & -1.7192 & 0.6716 & 0.0331 & 7.9\% \\
$ V $ & $ V-R_c $ &  21 & -0.01 & 1.23 & 0.5100 & 1.2159 & -0.0736 &  &  & 0.0191 & 4.5\% \\
$ V $ & $ V-I_c $ &  42 & -0.02 & 2.78 & 0.4992 & 0.6895 & -0.0657 &  &  & 0.0238 & 5.6\% \\
$ R_c $ & $ B-R_c $ &  21 & 0.00 & 2.80 & 0.4922 & 0.4933 & -0.1639 & 0.0486 &  & 0.0172 & 4.0\% \\
$ R_c $ & $ V-R_c $ &  21 & -0.01 & 1.23 & 0.5100 & 1.0159 & -0.0736 &  &  & 0.0191 & 4.5\% \\
$ R_c $ & $ R_c-I_c $ &  21 & -0.01 & 1.55 & 0.4989 & 1.2300 & -0.3549 &  &  & 0.0214 & 5.1\% \\
$ I_c $ & $ B-I_c $ &  42 & -0.01 & 4.35 & 0.4996 & 0.2424 & -0.0115 &  &  & 0.0213 & 5.0\% \\
$ I_c $ & $ V-I_c $ &  42 & -0.02 & 2.78 & 0.4992 & 0.4895 & -0.0657 &  &  & 0.0238 & 5.6\% \\
$ I_c $ & $ R-I_c $ &  21 & -0.01 & 1.55 & 0.4989 & 1.0300 & -0.3549 & 0.0291 &  & 0.0214 & 5.1\% \\
\hline
\end{tabular}
\end{table*}

\section{Discussion \label{discussion}}

\subsection{Reddening corrections and metallicity}

When a SBC relation is applied to a star suffering some amount of absorption, its colors must
be corrected for extinction before applying the relation. The amount of extinction is generally measured by the
color excess $E(B-V)$, and the conversion to other photometric bands depends on the intrinsic color of the star
and on the amount of extinction (Dean et al.~\cite{dean78}). For blue stars, the color excess
ratios are typically $E(V-R_c) = 0.60 \, E(B-V)$ and $E(V-I_c) = 1.25 \, E(B-V)$.
For red stars, the coefficients may increase by up to 10\%.

All the stars in our list are very nearby. In average, we do not expect a significant deviation of their
metallicity compared to that of the Sun. The exoplanet stars observed by Baines et al.~(\cite{baines08})
probably present a slight overmetallicity, but the effect on the dispersion of the adjusted SBC relations
is taken into account in the dispersions stated in Table~\ref{coefficients}. As discussed by
K04, the effect of metallicity on the visible-infrared SBC relations is
undetectable. For the visible SBC relations presented here, different metallicities will translate into
more significant photometric color differences. However, we expect them to be small compared
to the measured intrinsic dispersion of the relations, at least over a $\pm 0.5$\,dex range
around solar metallicity.

\subsection{Application to transits and microlensing}

The existence of exoplanetary transits in front of a star allows to retrieve its linear radius, as well as the
planetary radius. However, when the parallax is not known accurately and/or the star is too faint for high
accuracy angular velocity measurements (from high resolution spectroscopy), they are based on
an {\it a priori} radius for the star. The relations established in the present Research Note
are useful in this context, as they allow to retrieve the angular size of faint stars for which only broadband
photometry is available.
As a test, we can apply the $(V, V-I_c)$ relation to the nearby transiting exoplanet
stars \object{HD 189733}, for which $V = 7.686 \pm 0.007$, and $(V-I_c)=0.93 \pm 0.01$
(from Hipparcos). Due to the proximity of this star, we neglect the reddening.
For these bands, the surface brightness relation is (Table~\ref{coefficients}):
\begin{equation}
\log(\theta_{\rm LD}) = 0.4992 + 0.6895\,(V-I_c)  - 0.0657\,(V-I_c)^2 - 0.2\,V
\end{equation}
and we obtain $\theta_{\rm LD}({\rm HD 189733}) = 0.352 \pm 0.020$\,mas.
The attached 6\% uncertainty contains the photometric uncertainties and the intrinsic
dispersion of the $(V, V-I_c)$ relation. Together with the Hipparcos parallax
of this star $\pi = 51.94 \pm 0.87$\,mas, we obtain a linear photospheric radius of
$0.728 \pm 0.044\ R_\odot$.
This is in excellent agreement with the radius derived from $HST$ transit observations
by Pont et al.~(\cite{pont07}) of
$0.755 \pm 0.011\ R_\odot$.
We can also compare the $(V, V-I_c)$ angular diameter to the one computed
from the infrared $(V,V-K)$ linear relations calibrated by K04.
The 2MASS catalogue (Skrutskie et al.~\cite{skrutskie06}) lists $K =  5.541 \pm 0.021$,
giving $\theta_{\rm LD}({\rm HD 189733}) = 0.372 \pm 0.007$\,mas. Although this
value is within 1$\sigma$ of the $(V, V-I_c)$ value, the uncertainty on
this visible-infrared prediction is only one third of the visible version.
A complete review of the properties of this system can be found in
Torres, Winn \& Holman~(\cite{torres08}).

Another important application of SBC relations is related to gravitational microlensing studies, when
one needs to estimate the source radius from color-magnitude diagram (CMD) in the direction of the Galactic Bulge. 
Deredenned source magnitude and color are obtained from the relative position of the source in a CMD, 
with respect to the mean position of the Red Giant Clump (RGC).
As the intrinsic position of the RGC is calibrated from nearby stars by Hipparcos, the assumption
that the source suffers the same amount of extinction as the clump avoids the need for
an estimate of the extinction (see Yoo et al.~\cite{yoo04} for details). Generally, the magnitudes are
measured in the $I_c$ band, and the measured color is $(V-I_c)$, because this is what OGLE measures.
A caveat of K04's SBC relations is that they are attached to the Johnson $R$ and $I$ bands, and
this has been a source of confusion in the past.
Let's take the example of the recently discovered 3 Earth mass planet event MOA-2007-BLG-192
(Bennett et al.~\cite{bennett08}). The source appears to be fainter than the RGC by 5.7 mag in $I_c$,
and redder by 0.07\,mag in $(V-I_c)$. At the source distance (7.5\,kpc),
the RGC is at $I_{c0}({\rm RGC}) = 14.13$ and $(V-I_c)_0\,({\rm RGC}) = 1.04$,
so the source has $I_{c0} (\rm MOA\,192) = 19.84 \pm 0.24$ and $(V-I_c)_0\,(\rm MOA\,192) = 1.11 \pm 0.24$
(see Bennett et al.~\cite{bennett08} for a detailed discussion of these values).
Applying the appropriate $(I_c, V-I_c)$ SBC relation from Table~\ref{coefficients}
gives an estimated angular diameter of $\theta_{\rm LD}({\rm MOA\,192}) = 0.99 \pm 0.14\,\mu$as,
which is resolved by the caustic crossing (or cusp approach).

\subsection{Comparison with other calibrations}

In their early publication, Barnes et al.~(\cite{barnes78}) obtained a linear SBC relation
$F_V = 3.964 - 0.333\,(B-V)$ for $0.10 \leqslant (B-V) \leqslant 1.35$, with a typical
residual scatter of $\sigma(F_V) = \pm 0.025$. Following K04, it can be expressed as
$\log(\theta_{\rm LD}) =  0.5134 + 0.666\,(B-V) - 0.2\,V$
with a scatter of $\sigma(\log \theta_{\rm LD}) = \pm 0.05$, equivalent to $\sigma(\theta_{\rm LD}) = \pm 12\%$.
Over their common range of applicability, the relative difference $\rho$ in $\theta_{\rm LD}$
with our polynomial relation (coefficients listed in Table~\ref{coefficients}) is
\begin{equation}
\rho({\rm B78}) = \left[{\theta_{\rm LD}({\rm B78}) - \theta_{\rm LD}({\rm K08})}\right]/{\theta_{\rm LD}({\rm K08})} = 3.6 \pm 3.5\%.
\end{equation}
We can also compare our $(B,B-I)$ predictions (as an example) with the $(V,V-K)$ relations calibrated
recently by Beuermann~(\cite{beuermann06}, hereafter B06). The $S_V(V-K)$ relation of this author's
Table~2 can be reformulated as:
\begin{equation}
\log \theta_{\rm LD} = 0.524\,(V-K)^2 + 0.267\,(V-K) - 1.221\,10^{-3} - 0.2\,V
\end{equation}
For the 34 stars in our sample, we obtain an average value of the relative difference with our calibration
of $\rho({\rm B06}) = 4.1 \pm 5.2\%$. One should note that B06's calibration relies on several data sources,
including indirect angular diameter estimates.
The same comparison with K04's $(V, V-K)$ relation gives $\rho({\rm K04}) = 0.7 \pm 4.6\%$.
The present calibration of the visible SBC relations is therefore compatible with
previously determined SBC relations using visible and visible-IR color indices.

\section{Conclusion}

We computed polynomial SBC relations in $BVR_cI_c$ that can be used to predict
the angular size of individual stars for which only broadband photometry is available.
The visible-infrared linear relations established by K04
usually provide more accurate angular size estimates. However, in many cases the $JHK$ band 
magnitudes of field stars are not available (due to crowding in 2MASS for instance),
and the present $BVR_cI_c$ relations will provide a reliable 
photospheric angular diameter prediction with a typical uncertainty of $\approx 5$\%.
Their compatibility with visible-IR relations gives further
confidence in their accuracy.

\begin{acknowledgements}
This research made use of the SIMBAD and VIZIER databases at the CDS (Strasbourg, France),
the Hipparcos catalogues, the Lausanne General Catalogue of 
Photometric Data, the General Catalogue of Variable Stars, and NASA's ADS Bibliographic Services. 
We also received the support of PHASE, the high angular resolution partnership between 
ONERA, Observatoire de Paris, CNRS, and University Denis Diderot Paris 7.
\end{acknowledgements}

{}

\end{document}